\documentclass[twocolumn,prl,aps,showpacs,superscriptaddress]{revtex4}
\usepackage{epsf,latexsym,graphicx,booktabs}

\begin{document}

\def \ba {\begin{eqnarray}}
\def \ea {\end{eqnarray}}
\def \vk {\mathbf{k}}
\def \vq {\mathbf{q}}
\def \Bb {Bi$_2$Sr$_2$CaCu$_2$O$_{8+\delta}$}
\def \BP {(Bi,Pb)$_2$(Sr,La)$_2$CuO$_{6+\delta}$}
\def \LSCO {La$_{2-x}$Sr$_x$CuO$_4$}
\def \YBCO {YBa$_2$Cu$_3$O$_{6.6}$}

\title{Sharp low energy feature in single-particle spectra
due to forward scattering in $d$-wave cuprate superconductors}

\author{Seung Hwan Hong}\email[Present address: School of Computational Sciences, Korea
Institute for Advanced Study, Seoul 130-722, Korea.]{}
\affiliation{Department of Physics and Institute for Basic
Science Research, SungKyunKwan University, Suwon 440-746, Korea.}

\author{Jin Mo Bok}\email[Present address: National Laboratory for
Superconductivity, Institute of Physics, Chinese Academy of
Sciences, Beijing 100190, China]{} \affiliation{Department of
Physics and Institute for Basic Science Research, SungKyunKwan
University, Suwon 440-746, Korea.}

\author{Wentao Zhang}
\author{Junfeng He}
\author{X. J. Zhou}
\affiliation{National Laboratory for
Superconductivity, Beijing National Laboratory for Condensed
Matter Physics, Institute of Physics, Chinese Academy of
Sciences, Beijing 100190, China}

\author{C. M. Varma}
\affiliation{Department of Physics and Astronomy, University of
California, Riverside, California 92521. }

\author{Han-Yong Choi}\email[To whom the correspondences should
be addressed: ]{hychoi@skku.ac.kr. } \affiliation{Department of
Physics and Institute for Basic Science Research, SungKyunKwan
University, Suwon 440-746, Korea.} \affiliation{Asia Pacific
Center for Theoretical Physics, Pohang 790-784, Korea.}

\begin{abstract}

There is an enormous interest in renormalization of
quasi-particle (qp) dispersion relation of cuprate
superconductors both below and above the critical temperature
$T_c$ because it enables determination of the fluctuation
spectrum to which the qps are coupled. A remarkable discovery by
angle-resolved photoemission spectroscopy (ARPES) is a sharp low
energy feature (LEF) in qp spectra well below the superconducting
energy gap but with its energy increasing in proportion to $T_c$
and its intensity increasing sharply below $T_c$. This unexpected
feature needs to be reconciled with $d$-wave superconductivity.
Here, we present a quantitative analysis of ARPES data from \Bb \
(Bi2212) using Eliashberg equations to show that the qp
scattering rate due to the forward scattering impurities far from
the Cu-O planes is modified by the energy gap below $T_c$ and
shows up as the LEF. This is also a necessary step to analyze
ARPES data to reveal the spectrum of fluctuations promoting
superconductivity.

\end{abstract}



\maketitle

The accumulated high resolution ARPES data on the Bi-cuprate
(BSCCO) over a wide doping and temperature range have revealed a
sharp feature in the spectral function
\cite{Zhang08prl,Plumb10prl,Vishik10prl,Anzai10prl,Kondo13prl,Peng13cpl}.
This occurs below about 10 meV along the nodal $(0,0)-(\pi,\pi)$
direction in the Brillouin zone. Further investigations revealed
that the energy of the feature tracks $T_c$ regardless of the
families of BSCCO or doping concentration, that its position is
much smaller than the maximum energy gap $\Delta_0$, and that its
strength is sharply enhanced below $T_c$. Here, we show that all
these aspects of LEF can be explained by the forward scattering
impurities located off the Cu-O planes. We substantiate this by
combining computation of the self-energy using the Eliashberg
equations and analysis of the momentum distribution curve (MDC)
of the ultra high resolution laser based ARPES intensity from
Bi2212 in terms of the self-energy\cite{Zhang12prb,Yun11prb}. The
off-plane impurity parameters from fitting the LEF in the
superconducting (SC) state agree well with the normal state
scattering rate determined independently as discussed below. This
means that the qp scattering rate due to the off-plane impurities
is modified by the energy gap below $T_c$ and remarkably shows up
as the LEF in the $d$-wave SC state. The idea that forward
scattering in $d$-wave superconductors may lead to unusual
spectroscopic feature in SC state was already pointed out by Zhu,
Hirschfeld and Scalapino
\cite{Zhu04prb,Scalapino06jpcs,Dahm05prb,Graser07prb}. Here, we
quantitatively show the relevance of the idea by analyzing the
ARPES experiments at various angles and temperatures to producing
the low energy feature. We also compare the off-plane impurity
idea with the proposal that the LEF may be due to scattering from
acoustic phonons \cite{Johnston12prl}.

The qp scattering rate $\Gamma$ measured by ARPES in the normal
state depends on direction on the Fermi surface and is
significantly larger than $k_B T_c$, but this scattering does not
show up in the resistivity \cite{Valla99science}. This led to the
proposal that it is due to dopant impurities which lie in between
the Cu-O planes \cite{Abrahams00pnas}. Such impurities lead only
to small angle or forward scatterings for qps near the Fermi
surface. The characteristic scattering angle is $\delta\theta \sim
a/d$, where $a$ and $d$ are the in-plane and out-of-plane lattice
constants. Indeed were this not forward scattering, its effect
would exceed the Abrikosov-Gorkov bounds on impurity scattering
in $d$-wave superconductors and there would be no high $T_c$
superconductivity.

The ARPES intensity is given by
 \ba
 \label{intensity}
I(\vk,\omega)= |M(\vk,\nu)|^2 f(\omega)
\left[A(\vk,\omega)+B(\vk,\omega)\right],
 \ea
where $M$ is the matrix element, $\nu$ the energy of incident
photon, $f$ the Fermi distribution function, $A= -\frac1\pi
Im\left[ G\right]$ is the spectral function, and $B$ is the
background from the scattering of the photo-electrons. $G$ is the
Green's function,
 \ba \label{greens}
G(\mathbf{k},\omega)= \frac{W+Y} {W^{2}-Y^{2}-\phi^{2}},
 \ea
where $W= \omega-\widetilde{\Sigma}(\theta,\omega)$, $Y=
\xi(\mathbf{k})+X(\theta,\omega)$. $ \Sigma(\theta,\omega) =
\widetilde{\Sigma}(\theta,\omega) + X(\theta,\omega)$ is the
normal (or, diagonal) self-energy, $\phi(\theta,\omega)$ the
anomalous (off-diagonal) self-energy, and $\xi(\vk)$ is the bare
dispersion \cite{Sandvik04prb,Yun11prb}. We took
 \ba \label{bare}
\xi(\vk) = -t[\cos(k_x a)+\cos(k_y a)]-4t' \cos(k_x a)\cos(k_y a)
\nonumber \\ -2t''[\cos(2k_x a)+\cos(2k_y a)]-\mu,
 \ea
with $t=0.395$, $t'=0.084$, $t''=0.042$ eV, and $\mu=-0.43$
(UD89K) or $-0.48$ eV (OD82K). $\omega$ stands for the qp energy
with respect to the Fermi energy. The in-plane momentum $\vk$ is
written with the distance from the $(\pi,\pi)$ point, $k_\perp$,
and the tilt angle measured from the nodal cut, $\theta$. The MDC
analysis fits the measured ARPES intensity using Eqs.\
(\ref{intensity}) and (\ref{greens}) as a function of $k_{\perp}$
at fixed $\theta$ and $\omega$ to extract the
$k_\perp$-independent diagonal and off-diagonal self-energies,
$\Sigma(\theta,\omega)$ and $\phi(\theta,\omega)$. We note that
the dispersion relation from the Lorentzian MDC fitting is not
suitable for analyzing off-nodal cuts in SC state because the
ARPES dispersion bends back at $\omega\approx -\Delta(\theta)$
due to the gap and meaning of the dispersion relation becomes obscure below $\Delta(\theta)$.


Ultra high resolution laser ARPES data were collected from
slightly underdoped Bi2212 of the critical temperature $T_c=89$ K
and pseudogap temperature $T^*\approx 160$ K (denoted by UD89K)
and overdoped $T_c=82$ K Bi2212 samples (OD82K). The experimental
setup is the same as in the Ref.\ \onlinecite{Zhang12prb,He13prl}.
The raw ARPES intensity data from UD89K and OD82K together with
the extracted self-energy are presented in the supplemental
material (SM) [url], which includes Refs.\
\cite{He13prl,Matsui03prl}.

The qp self-energy may be expressed in terms of two terms, due to
coupling to a boson spectrum and to impurities. See below for
details. The self-energy from impurity scattering may be written
as
 \ba \label{self-energy_imp}
 \Sigma_{imp}(\mathbf{k},\omega)=
n_{imp}\sum_{\mathbf{k'}} \Big| V_{imp}(\mathbf{k},\mathbf{k'})
\Big|^{2} G(\mathbf{k'},\omega),
 \ea
where $n_{imp}$ and $V_{imp}$ are the impurity concentration and
impurity potential, respectively. $G$ is the retarded Green's function of Eq.\ (\ref{greens})
which includes the impurity effects as given by Eq.\ (\ref{impurity_cal}) below.
$\phi$ may also be decomposed similarly with
 \ba
\label{phi_imp} \phi_{imp}(\mathbf{k},\omega)= -
n_{imp}\sum_{\mathbf{k'}} \Big| V_{imp}(\mathbf{k},\mathbf{k'})
\Big|^{2} \frac{\phi(\mathbf{k'},\omega)} {W^{2}-Y^{2}-\phi^{2}}.
 \ea

For a momentum independent scattering potential, the self-energy
of Eq.\ (\ref{self-energy_imp}) at ${\bf k} = {\bf k}_F$ reduces
after integrating over $\vk'$ to $ \Sigma_{imp}(\omega)= -i\Gamma
\left<N(\theta',\omega)\right>_{\theta'},$ where $
N(\theta,\omega) = {\omega}/\sqrt{\omega^2-\Delta_0^2
\sin^2(2\theta)}.$ After the average over angle $\left<
\right>_{\theta'}$, $\Sigma_{imp}(\omega)$ has an angle
independent peak at $\omega=\Delta_0$ below $T_c$, and it reduces
to $\Sigma_{imp} = -i \Gamma$ above $T_c$ as expected. For the
strong forward scattering limit of $\theta'\approx\theta$, by
contrast,
 \ba \label{model_forward}
\Sigma_{imp}(\theta,\omega)= -i\Gamma(\theta) N(\theta,\omega),
 \ea
where $\Gamma(\theta) =\pi n_{imp}V_{imp}^2 N_F(\theta)$ and
$N_F(\theta)\sim 1/v_F(\theta)$ is the angle dependent DOS at the
Fermi surface, so that the impurity self-energy depends on the
direction $\theta$.

\begin{figure}[tbh]
\includegraphics[width=9cm]{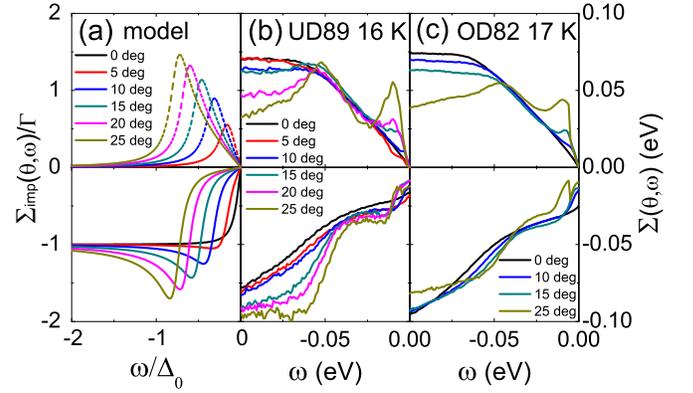}
\caption{(a) The impurity self-energy from the model calculation
of Eq.\ (\ref{model_forward}). The real and imaginary parts are
shown along several cuts with the dashed and solid curves,
respectively. (b)-(c) The extracted self-energy from MDC analysis
of UD89K at $T=$ 16 K and of OD82K at $T=17$ K. }
 \label{fig:model-self}
\end{figure}

In Fig.\ \ref{fig:model-self}(a) we plot the model self-energy of
Eq.\ (\ref{model_forward}) with a small imaginary part $\Delta_2 =
0.1 \Delta_0$. Notice a kink in the self-energy whose position
and strength increase as the tilt angle increases. For
comparison, we show in Fig.\ \ref{fig:model-self}(b) and (c) the
extracted qp self-energy from the MDC analysis of UD89K and OD82K
data in SC state. The LEF shows up below $|\omega| \lesssim 20$
meV. The higher, $\gtrsim$ 50 meV, feature is from inelastic
coupling to boson fluctuations and will be further discussed
below. In agreement with the model calculation, the LEF position
in Fig.\ \ref{fig:model-self}(b) and (c) increases as the tilt
angle increases and is smaller than the maximum gap $\Delta_0
\approx 20$ meV. Recall that the peak position in
$-\Sigma_2(\vk,\omega)$ from coupling to a boson mode of a peak
at $\omega_b$ (with a broad momentum dependence) is given by
$\omega_b +\Delta_0$.\cite{Sandvik04prb} The observed feature is
impossible to understand with this picture because the observed
peak position is less than $\Delta_0$ and angle dependent. But,
this feature is naturally understood from the forward scattering
off-plane impurities, just as in the normal state.

\begin{figure}[tbh]
\includegraphics[width=9cm]{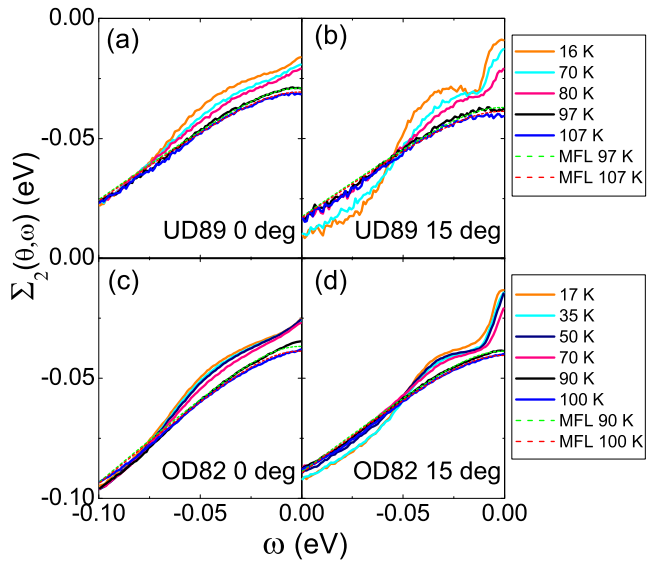}
\caption{The imaginary parts of the extracted self-energy from
UD89K and OD82K data both above and below $T_c $ along
$\theta=0^\circ$ and $15^\circ$. The dashed curves are the
fitting from Eq.\ (\ref{mfl}). Notice that the LEF is sharply
enhanced below $T_c$ especially along off-nodal cuts. }
\label{fig:imp_self2}
\end{figure}

Now, we consider how the self-energy changes as $T$ varies for a
given cut $\theta$. In Fig.\ \ref{fig:imp_self2}, we show the
imaginary parts of the extracted self-energy from UD89K and OD82K
along $\theta=0$ and $\theta=15^\circ$. The LEF emerges sharply
below $T_c$ and the kink energy increases as $T$ is reduced.
Although the LEF is much weaker along the nodal cut compared with
off-nodal cuts, we show the nodal cut results to make more
concrete contacts with the published
reports.\cite{Zhang08prl,Plumb10prl,Kondo13prl,Peng13cpl} For
instance, compare Fig.\ \ref{fig:imp_self2}(a) and (c) with the
inset of Fig.\ 4(a) in Ref.\ \onlinecite{Peng13cpl} and notice
their similarity. In the normal state it is well established that
the ARPES measured scattering rate may be represented as a sum of a constant term
and a frequency dependent one as
 \ba\label{mfl}
-\Sigma_2(\theta,\omega) = \Gamma(\theta)+ b(\theta)
\sqrt{\omega^2 +(\pi T)^2 },
 \ea
where the constant $\Gamma$ is from the off-plane impurities as discussed in
the introduction.\cite{Abrahams00pnas,Kaminski05prb} The frequency dependence is
well represented by the marginal Fermi liquid (MFL) form\cite{Varma89prl} as noted
previously.\cite{Valla99science,Abrahams00pnas,Kaminski05prb} The dashed
curves in Fig.\ \ref{fig:imp_self2} represent the fitting to Eq.\
(\ref{mfl}). The obtained parameters from the normal state fitting are
discussed below in comparison with SC state consideration.

\begin{figure}[tbh]
\includegraphics[width=9cm]{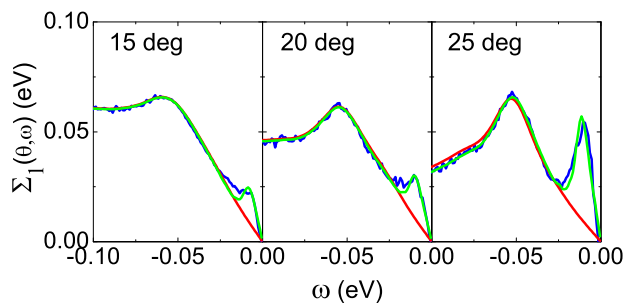}
\caption{ The real parts of the self-energy of UD89K at 16 K. The
blue and green curves represent the extracted and calculated (Eq.\
(\ref{self_eff2})) self-energy, and the red curves show the
calculated self-energy with the impurity term removed. }
 \label{fig:f0remove}
\end{figure}

We now put the above discussion in a quantitative basis. First,
we wish to establish that the LEF mandates the emergence of the
zero frequency mode. Because the zero frequency mode represents
the impurity term as discussed below, it formally implies that the
impurities generate the LEF. This is done by an inversion of the
Eliashberg equations. As shown in the SM, the impurity
self-energy may be included in the qp self-energy as the static
component of the Eliashberg function $\alpha^2
F^{(+)}(\epsilon'=0)$ as
 \ba \label{self_eff2}
\Sigma(\theta,\omega) = \int^{\infty}_{-\infty} d\epsilon'
M(\omega,\epsilon') \alpha^2 F^{(+)}(\theta,\epsilon'),
 \\ \label{kernel}
M(\omega,\epsilon')= \int^{\infty}_{-\infty} d\epsilon
\frac{f(\epsilon)+n(-\epsilon')}{\epsilon+\epsilon'-\omega-i\delta}
\left< Re N(\theta',\epsilon) \right>_{\theta'},
 \ea
where $n$ is the Bose distribution function. Using the MDC
extracted $\Sigma(\theta,\omega)$ as an input for the left hand
side of the Eq.\ (\ref{self_eff2}), we may invert this equation to
obtain the $\alpha^2 F^{(+)}(\theta,\epsilon')$. The $\alpha^2 F
$ so obtained is the fluctuation spectrum for $\epsilon'\ne 0$,
and the impurity scattering for $\epsilon'= 0$. The inversion was
performed using the maximum entropy method (MEM) as reported
previously.\cite{Yun11prb}

We present in Fig.\ \ref{fig:f0remove}, the  extracted
$\Sigma_1(\theta,\omega)$ with the blue curves and the calculated
self-energy from the MEM inversion of Eq.\ (\ref{self_eff2}) with
the green curves. The strong forward scattering limit was used in
the angle averaged DOS in the $M(\omega,\epsilon')$ for
$\epsilon'=0$ of Eq.\ (\ref{kernel}) in the inversion. Then, we
removed $\alpha^2 F^{(+)}(\theta,\epsilon'=0)$ and recalculated
the self-energy. The results are shown with the red curves where
the LEF is conspicuously absent. This formally establishes that
the LEF is due to the forward scattering off-plane impurities. The
slight misfit only indicates that the actual impurities in the
Bi2212 scatter the electrons more broadly than the extreme
forward scattering limit as we show by calculations below.

\begin{figure}[tbh]
\includegraphics[width=9cm]{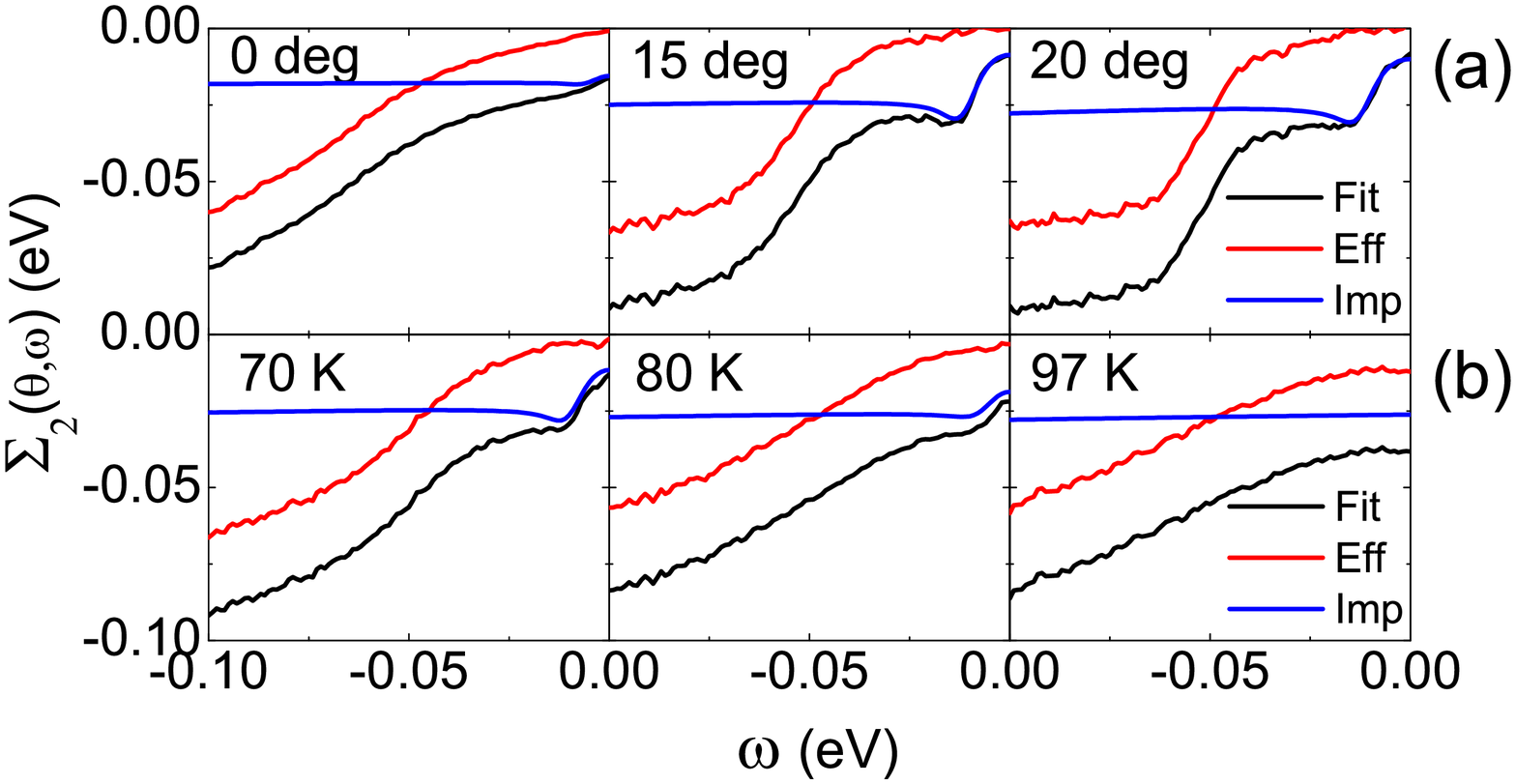}
\caption{ Comparison of the imaginary parts of the calculated
impurity self-energy (blue curves) with the extracted one (black)
for UD89K. The red curves represent their difference, that is,
the intrinsic self-energy due to coupling to bosons. Figure (a)
is as the angle varies at $T=16$ K, and (b) is as the temperature
varies at $\theta=15^\circ$.} \label{fig:vimp}
\end{figure}

We now present a second quantitative support for the out-of-plane
impurity idea for the LEF with a specific impurity potential. A
reasonable model is \cite{Zhu04prb}
 \ba \label{V_imp}
V_{imp}(\mathbf{k},\mathbf{k'}) = \frac{2\pi\kappa V_{0}}{\left[
(\mathbf{k}-\mathbf{k'})^{2}+\kappa^{2} \right]^{3/2}}.
 \ea
$\kappa^{-1}$ is the range of the impurity potential which
controls the angle dependence of the impurity self-energy. We
calculated
 \ba\label{impurity_cal}
\Sigma(\vk,\omega) = \Sigma_{imp}(\vk,\omega) -i \Gamma_0 , \nonumber \\
\phi(\vk,\omega) = \phi_{imp}(\vk,\omega) +\phi_0 [\cos(k_x
a)-\cos(k_y a)]/2,
 \ea
where the parameters are to be determined by fitting the
self-energy quantitatively. Eq.\ (\ref{impurity_cal}) was solved
self-consistently together with Eqs.\ (\ref{self-energy_imp}) and
(\ref{phi_imp}). We compare in Fig.\ \ref{fig:vimp} the imaginary
parts of the calculated impurity self-energy against that
extracted from experiments in UD89K. Fig.\ \ref{fig:vimp}(a)
shows the self-energy as the tilt angle varies at $T=16$ K, and
(b) the temperature evolution at $\theta=15^\circ$ at $T=70$, 80,
and 97 K. The comparison of the real part of the self-energy and
those from OD82K is presented in SM for completeness. The
observed LEF are well reproduced by the off-plane impurity
calculations. We stress that the prominent LEF in the self-energy
in SC state enables one to accurately determine the parameters of
the impurity potential, just like the electron-phonon interaction
function can be most accurately determined in SC state. Of the
low energy peak, $\phi_0$ determines the energy scale, $n_{imp}
V_0^2$ the magnitude, $\kappa$ the angle dependence of the
position, and $\Gamma_0$ sets the width. $\Gamma_0$ includes the
effects of finite temperature and disorder other than the
off-plane impurities.

As shown in Fig.\ \ref{fig:vimp}, we get very good fits to the
data. The change of the parameters required in going from UD89K to
OD82K is that $\phi_0$ decreases from 19 to 16 meV and
$n_{imp}V_{0}^2$ (angle average) from 0.031 to 0.038 $t^2 a
\kappa^3$, with $t$ the nearest neighbor hopping amplitude. Other
parameters remain almost the same; $\kappa \approx 0.3/a$ and
$\Gamma_0 \approx 3$ meV. This is reasonable because $\kappa\sim
k_F \delta\theta \sim (1/a)(a/d)$ for off-plane impurity
scattering as discussed in the introduction and is expected to be
insensitive to doping. But $n_{imp}V_0^2$ is expected to increase
with doping and is indeed found larger for OD82K.

The normal state scattering rate from the impurity potential of
Eq.\ (\ref{V_imp}) in the strong forward scattering limit is
given by \cite{Zhu04prb}
 \ba \label{gamma_zhu}
\Gamma_V (\theta)= -Im \Sigma(\theta,\omega=0) =
 \frac{3\pi^2 n_{imp} V_0^2 }{8 v_F(\theta) \kappa^3 }.
 \ea
We calculated $\Gamma_V $ to check consistency of the idea, using
the parameters from deep in SC state (and $\hbar v_F\approx
3.9-2.8$ eV\AA \ for $\theta=0-20^\circ$ from the bare dispersion
$\xi(\vk)$, and the lattice constant $a \approx 3.8$ \AA). They
are in a good agreement with the $\Gamma(\theta)$ of Eq.\
(\ref{mfl}) from the independent normal state fitting. They are
tabulated in detail with other parameters in the Table 1 in the
SM. This confirms the idea that the $\theta$-dependent qp
scattering rate due to the forward scattering impurities is
modified by the energy gap below $T_c$ and shows up as the LEF.

Recently, Johnston $et~al.$ suggested that the LEF is caused by
the acoustic phonons which make forward scatterings due to poor
metallicity.\cite{Johnston12prl} $\kappa$ in this scenario is set
by the Thomas-Fermi wavevector and should increase with doping
because of better metallicity. This seems at odds with what is
found here. The doping dependence of the LEF has been studied by
several groups.\cite{Anzai10prl,Kondo13prl,Peng13cpl} Ref.\
\onlinecite{Anzai10prl,Kondo13prl} reported that the LEF is
enhanced as doping is reduced. But, the recent study found that
the low energy kink position becomes reduced in the heavily
underdoped samples, which seems inconsistent with the acoustic
phonon picture.\cite{Peng13cpl} The kink energy in the off-plane
impurity scenario is set by the gap and its decrease in the
heavily underdoping regime is naturally understood. So is the
sharp change of the LEF across $T_c$ because it is given directly
by the SC DOS. However, LEF in the acoustic phonon picture is
expected to be smoother because it is given by a convolution of
DOS and boson spectrum and because it should be present above as
well as below $T_c$ since SC is not prerequisite for this effect.
Compare the temperature evolution of the boson feature near
$50-100$ meV and the feature near $\sim$10 meV of the black
curves in Fig.\ \ref{fig:vimp}(b) and in Fig.\ 2(b) and (d), and
notice the $T$ evolution of the $50-100$ and $\sim$10 meV
features is in sharp contrast.

In summary, we have proposed here that the sharp low energy
feature observed below $\sim$10 meV in BSCCO is indeed caused by
the forward scattering off-plane impurities. This conclusion is
based on the following observations: (1) the LEF mandates
emergence of the Eliashberg spectrum at zero frequency as shown
in Fig.\ \ref{fig:f0remove}, (2) the impurity potential well
produces the sharp LEF as the angle or temperature is varied as
shown in Fig.\ \ref{fig:vimp}, (3) the parameters of the impurity
potential obtained from the SC state satisfactorily match the
normal state scattering rates from independent determination, (4)
the change of parameters between UD89K and OD82K is consistent
with the off-plane impurity idea.

It should be clear by now that the low energy feature of BSCCO
ARPES data is from the forward scattering off the out-of-plane
impurities. After removing this impurity part from the MDC
extracted self-energy we can uncover the intrinsic self-energy as
shown by the red curves in Figs.\ 3 and 4. It is this impurity
removed self-energy which must be used as an input to invert the
Eliashberg equation. This is a necessary step to reveal the boson
spectrum promoting high $T_c$ superconductivity.

{\it Acknowledgments} -- This work was supported by National
Research Foundation (NRF) of Korea through Grant No.\
NRF-2013R1A1A2061704. XJZ thanks financial support from the NSFC
(11190022,11334010 and 11374335) and the MOST of China (973
program No: 2011CB921703 and 2011CBA00110). CMV's work was
supported by NSF-DMR-1206298.

\bibliographystyle{apsrev}

\bibliography{review}

\end{document}


\def \ba {\begin{eqnarray}}
\def \ea {\end{eqnarray}}
\def \vk {\mathbf{k}}
\def \vq {\mathbf{q}}
\def \Bb {Bi$_2$Sr$_2$CaCu$_2$O$_{8+\delta}$}
\def \BP {(Bi,Pb)$_2$(Sr,La)$_2$CuO$_{6+\delta}$}
\def \LSCO {La$_{2-x}$Sr$_x$CuO$_4$}
\def \YBCO {YBa$_2$Cu$_3$O$_{6.6}$}

\title{ {\it Supplemental Material for} \\
Sharp low energy feature in single-particle spectra
due to forward scattering in $d$-wave cuprate superconductors}

\author{Seung Hwan Hong}\email[Present address: School of Computational Sciences, Korea
Institute for Advanced Study, Seoul 130-722, Korea.]{}
\affiliation{Department of Physics and Institute for Basic
Science Research, SungKyunKwan University, Suwon 440-746, Korea.}

\author{Jin Mo Bok}\email[Present address: National Laboratory for
Superconductivity, Institute of Physics, Chinese Academy of
Sciences, Beijing 100190, China]{} \affiliation{Department of
Physics and Institute for Basic Science Research, SungKyunKwan
University, Suwon 440-746, Korea.}


\author{Wentao Zhang}
\author{Junfeng He}
\author{X. J. Zhou}
\affiliation{National Laboratory for
Superconductivity, Beijing National Laboratory for Condensed
Matter Physics, Institute of Physics, Chinese Academy of
Sciences, Beijing 100190, China}

\author{C. M. Varma}
\affiliation{Department of Physics and Astronomy, University of
California, Riverside, California 92521. }

\author{Han-Yong Choi}\email[To whom the correspondences should
be addressed: ]{hychoi@skku.ac.kr. } \affiliation{Department of
Physics and Institute for Basic Science Research, SungKyunKwan
University, Suwon 440-746, Korea.} \affiliation{Asia Pacific
Center for Theoretical Physics, Pohang 790-784, Korea.}



\maketitle

\vspace{1cm}

{\bf ARPES intensity and MDC self-energy analysis \\}

The ARPES measurements were performed on the vacuum ultraviolet
(VUV) laser-based angle-resolved photoemission system with
advantages of high photon flux, enhanced bulk sensitivity, and
super-high energy and momentum resolution. The photon energy is
6.994 eV with a bandwidth of 0.26 meV. We set the energy
resolution of the electron energy analyzer (Scienta R4000) at 1
meV, giving rise to an overall energy resolution of 1.03 meV. The
angular resolution is $\sim$0.3$^\circ$, corresponding to a
momentum resolution of $\sim$0.004 \AA$^{-1}$ for the 6.994 eV
photon energy. The experimental setup is the same as the Ref.\
\onlinecite{Zhang12prb,He13prl}. The ultra high resolution laser
ARPES data were collected from slightly underdoped Bi2212 of the
critical temperature $T_c=89$ K and pseudogap temperature
$T^*\approx 160$ K (denoted by UD89K) and overdoped $T_c=82$ K
Bi2212 samples (OD82K).

We present in Fig.\ 1 the raw ARPES intensity from UD89K at $T=16
$ K in the first row and OD82K at $T=17$ K in the second row. In
Fig.\ 2 we show the dispersion relation and scattering rate
(a1-a6 for UD89K and c1-c6 for OD82K) and the real and imaginary
parts of the self-energy (b1-b6 for UD89K and d1-d6 for OD82K).
The dispersion relation and scattering rate were obtained from
the Lorentzian fitting of the momentum distribution curves (MDC),
and the self-energy from the MDC fitting using the full SC Green's
function as explained in the main text. The discrepancy between a
and b and between c and d comes from the fact that the
tight-binding dispersion and the full SC Green's function were
used in the Green's function fitting. The Lorentzian fitting
amount to using linear dispersion. The difference between $a1$
and $b1$ along the nodal cut arises only because the
tight-binding bare dispersion was employed in the self-energy
analysis.

As the tilt angle increases away from the nodal direction, the
shallower band bottom can be described using the tight-binding
dispersion which however is entirely missed by the linear
dispersion. Also, the ARPES dispersion bends back at
$\omega\approx -\Delta(\theta)$ due to pairing
\cite{Matsui03prl,Zhang12prb} and meaning of the dispersion
relation is not clear below the gap energy, which becomes
pronounced for large tilt angle. As Zhu $et~al.$ pointed out, the
near cancellation of the sharp features in the diagonal and
off-diagonal self-energies substantially weakens the feature in
the scattering rate.\cite{Zhu04prb} For instance, compare a6 and
b6 of UD89K at $\theta=25^\circ$ and $T=16$ K. The dispersion
relation and scattering rate in a6 show qualitatively different
behavior below $\sim$20 meV from the self-energy in b6. They can
be misleading as explained above. On the other hand, the
self-energy is well defined for all $\theta$ and $\omega$, and we
will discuss the LEF in terms of the self-energy from the MDC
analysis using the full SC Green's function. The LEF shows up
more clearly along off-nodal cuts as can be seen from the plots
in Fig.\ 1 and 2, and should be analyzed there properly.

\begin{figure*}[tbh]
\includegraphics[width=15.cm]{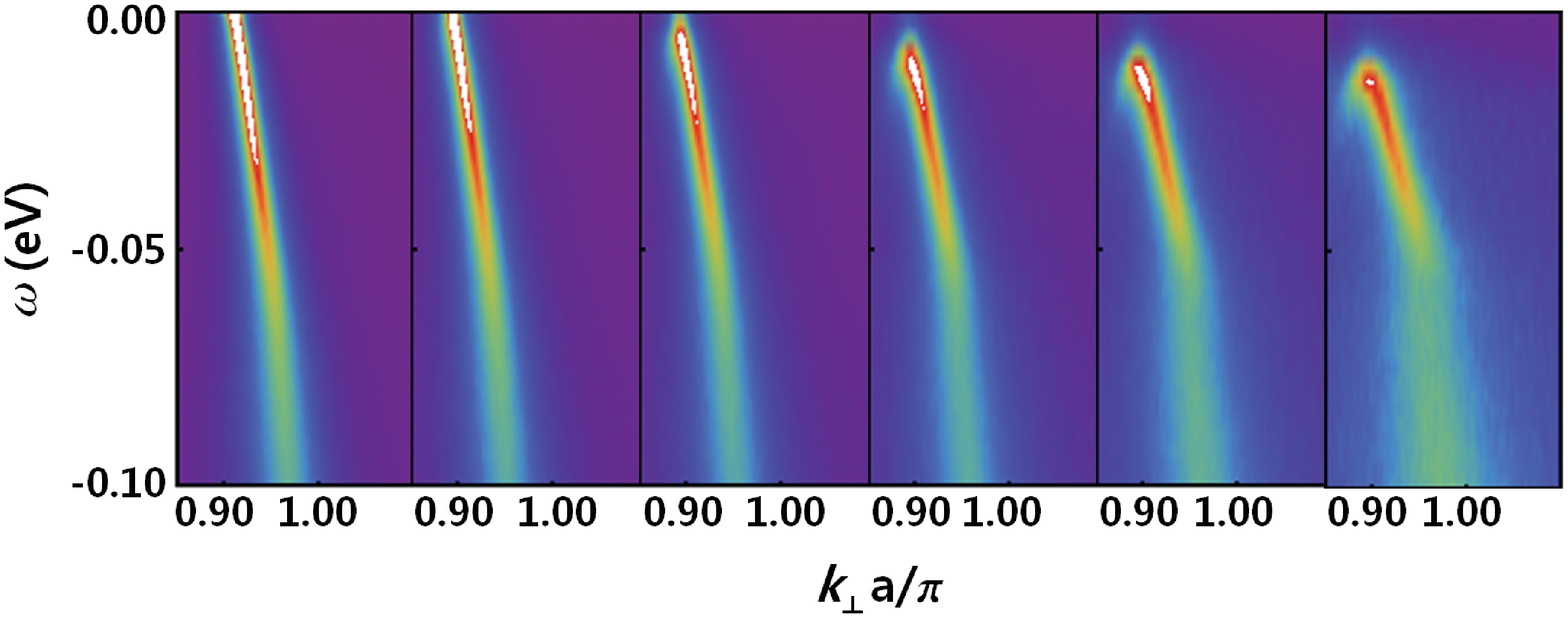}
\includegraphics[width=15.cm]{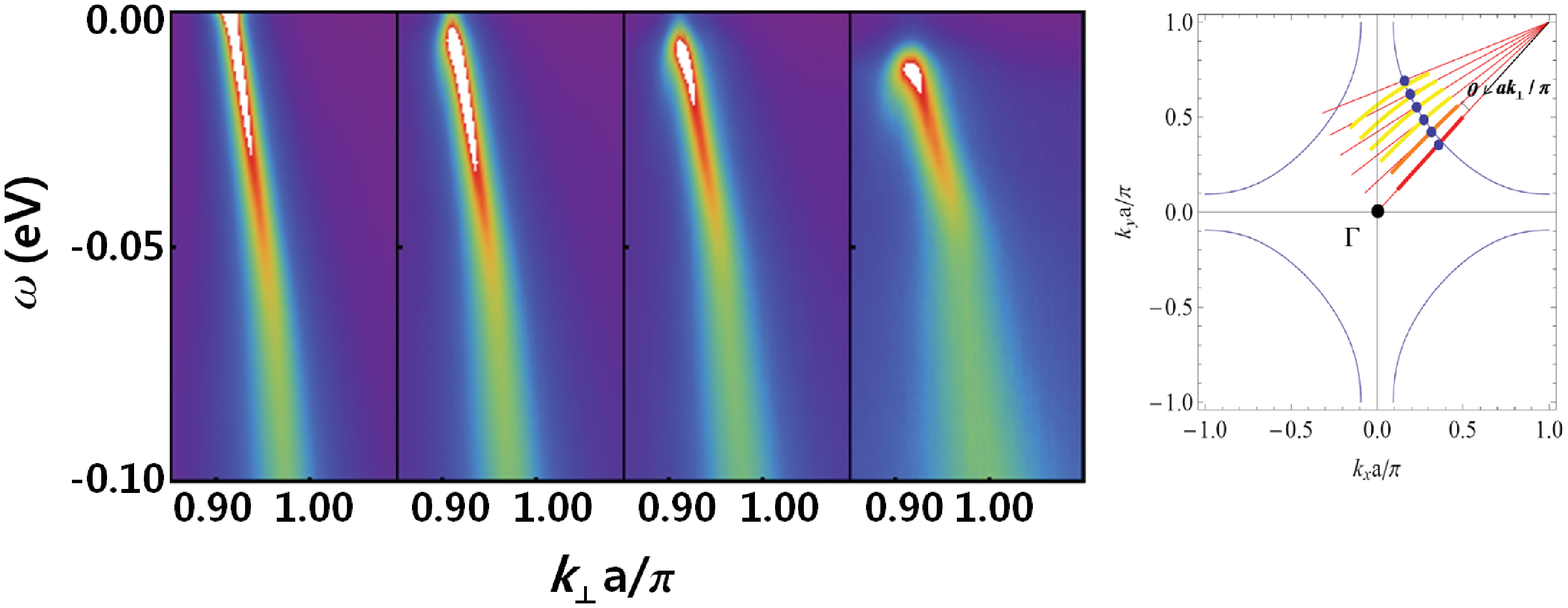}
\caption{The raw ARPES intensity from the UD89K and OD82K. The
first and second rows are the intensity from UD89K at $T=16$ K
and OD82K at $T=17$ K, respectively. The tilt angle for the first
row is 0, 5, 10, 15, 20, and 25 deg from the left, and for the
second row, the angle is 0, 10, 15, 25 deg. In the right corner,
the Fermi surface of Bi2212 is illustrated where $ k_\perp$ is the
distance from the $(\pi,\pi)$ point in the Brillouin zone and
$\theta$ is the tilt angle from the $(0,0)-(\pi,\pi)$ nodal
direction. }
 \label{fig:intensity}
\end{figure*}

\begin{figure*}
\includegraphics*[width=14cm]{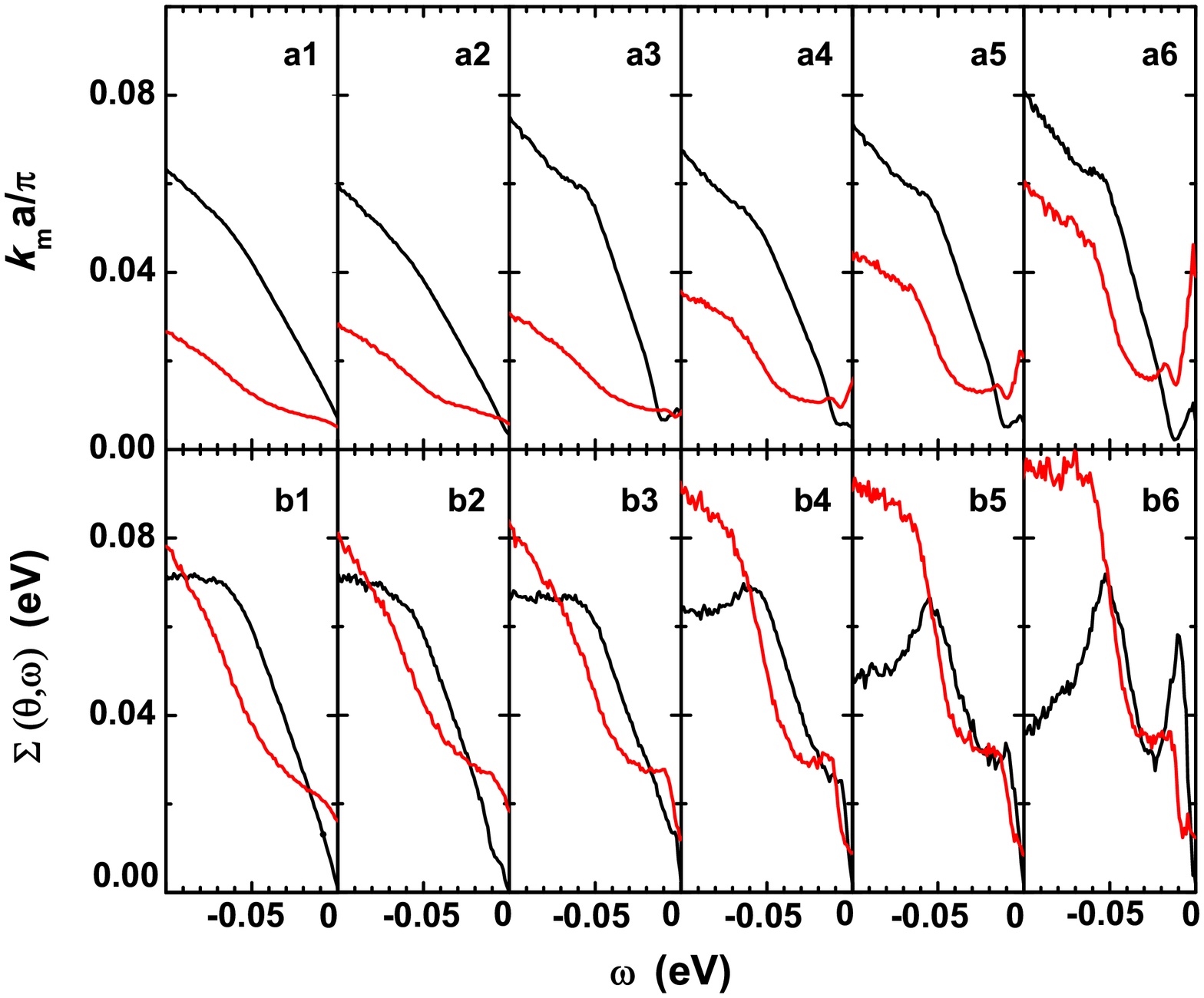}
\includegraphics*[width=14cm]{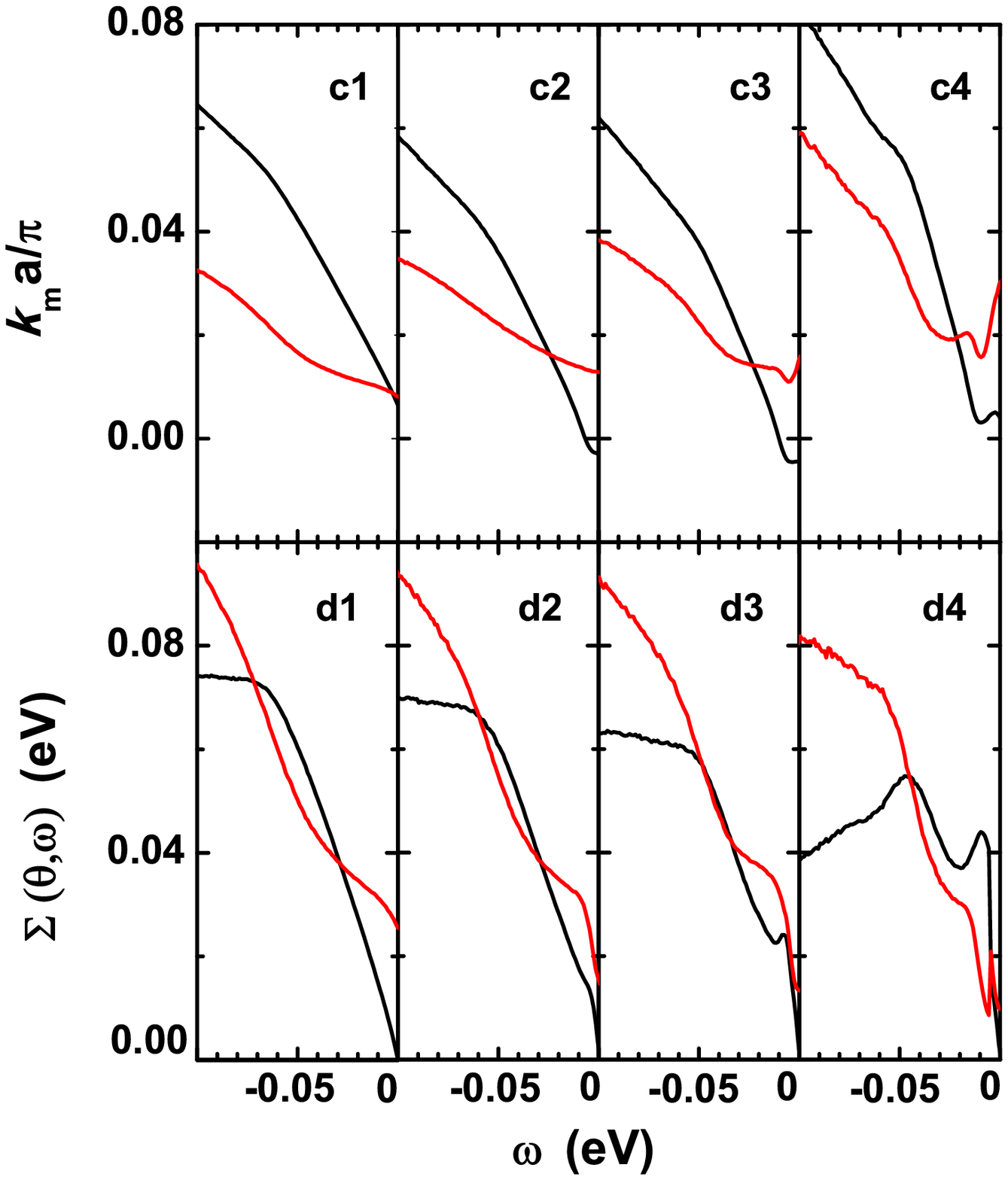}
\caption{ (a) \& (b) Analysis of the ARPES intensity of UD89K
shown in Fig.\ 1. (a) The dispersion relation and scattering rate
(HWHM) determined from the Lorentzian MDC fitting, shown in the
black and red, respectively. $k_m=k_\perp^{max}-k_0$, where
$k_\perp^{max}$ is the distance from the $(\pi,\pi)$ to the
maximum point where the MDC has a peak in the BZ and $k_0$ is the
distance from $(\pi,\pi)$ to the Fermi surface along a given
angle. The tilt angles are the same order as in Fig.\ 1. (b) The
real and imaginary parts of the self-energy determined from the
MDC fitting with the full SC Green's function as explained in the
main text. The black and red show the real part $\Sigma_1
(\theta,\omega)$ and minus of the imaginary part $-\Sigma_2
(\theta,\omega)$ as a function of $\omega$. (c) \& (d) Analysis
of OD82K intensity of Fig.\ 1. (c) The dispersion relation and
scattering rate of OD82K at $T=17$ K along $\theta=0$, 10, 15,
and 25$^\circ$ from left. (d) The $\Sigma_1 (\theta,\omega)$ in
black and $-\Sigma_2(\theta,\omega)$ in red of OD82K at $T=17$ K.
The small spiky feature of the red curve in d4 near $\sim$5 meV is
from a numerical instability and should be disregarded. }
 \label{fig:dispersion}
\end{figure*}

\newpage

\vspace{5cm}

{  \bf Eliashberg theory with boson and impurity \\ }

The qp self-energy may be decomposed into two contributions as
 \ba \label{self-energy}
\Sigma(\mathbf{k},\omega) = \Sigma_{eff}(\mathbf{k},\omega) +
\Sigma_{imp}(\mathbf{k},\omega).
 \ea
The first and second parts come from coupling to a boson spectrum
and to impurities, respectively, and may be written as
 \ba \label{self-energy_eff}
 \Sigma_{eff}(\mathbf{k},\omega) &=&
\int^{\infty}_{-\infty} d\varepsilon\int^{\infty}_{-\infty}
d\varepsilon'
\frac{f(\varepsilon)+n(-\varepsilon')}{\varepsilon+\varepsilon'-\omega-i\delta}
 \sum_{\mathbf{k'}} A(\mathbf{k'},\varepsilon)
\alpha^{2}F^{(+)}(\mathbf{k,k'},\varepsilon'), \\
 \label{self-energy_imp} \Sigma_{imp}(\mathbf{k},\omega) &=&
n_{imp}\sum_{\mathbf{k'}} \Big| V_{imp}(\mathbf{k},\mathbf{k'})
\Big|^{2} G(\mathbf{k'},\omega),
 \ea
where $f$ and $n$ are the Fermi and Bose distribution functions,
respectively, $\delta$ an positive infinitesimal number, and
$\alpha^2 F^{(+)}$ is the diagonal Eliashberg function. The
$n_{imp}$ and $V_{imp}$ are the impurity concentration and
impurity potential energy, respectively. The anomalous
self-energy $\phi$ may also be decomposed similarly as
\cite{Zhu04prb}
 \ba \label{phi}
\Sigma(\mathbf{k},\omega) = \Sigma_{eff}(\mathbf{k},\omega) +
\Sigma_{imp}(\mathbf{k},\omega),
 \ea
where the first and second terms are from the coupling to the
boson and impurities, as before, and may be written as
 \ba
\label{phi_eff}
 \phi_{eff}(\mathbf{k},\omega) &=& -
\int^{\infty}_{-\infty} d\varepsilon\int^{\infty}_{-\infty}
d\varepsilon'
\frac{f(\varepsilon)+n(-\varepsilon')}{\varepsilon+\varepsilon'-\omega-i\delta}
 \sum_{\mathbf{k'}} A_\phi(\mathbf{k'},\varepsilon)
\alpha^{2}F^{(-)}(\mathbf{k,k'},\varepsilon'), \\
 \label{phi_imp} \phi_{imp}(\mathbf{k},\omega) &=& -
n_{imp}\sum_{\mathbf{k'}} \Big| V_{imp}(\mathbf{k},\mathbf{k'})
\Big|^{2} G_\phi(\mathbf{k'},\omega),
 \ea
where $\alpha^2 F^{(-)}$ is the off-diagonal Eliashberg function,
and
 \ba \label{spectral_phi}
A_\phi(\mathbf{k},\omega)= -\frac1\pi Im\left[
G_\phi(\mathbf{k},\omega)\right], ~~ G_\phi(\mathbf{k},\omega)=
\frac{\phi(\mathbf{k},\omega)} {W^{2}-Y^{2}-\phi^{2}}.
 \ea

Now, by the symmetry requirement of
 \ba \label{a2fsymm} \alpha^2 F(-\epsilon')=-\alpha^2
F(\epsilon'),
 \ea
we have $\alpha^2 F(\epsilon'=0)=0$. But, the $n(\epsilon' )$ in
Eq.\ (\ref{self-energy_eff}) diverges as $\epsilon'\rightarrow
0$, and the product $\alpha^2 F(\epsilon')n(\epsilon')$ can be
finite, which may represent the impurity term as follows. Use
$n(-\epsilon')=-[n(\epsilon')+1]$, Eq.\ (\ref{a2fsymm}), and
 \ba
F(\omega)=\frac1\pi \int_{-\infty}^{\infty}d\epsilon
\frac{1}{\epsilon-\omega-i\delta} Im F(\epsilon),
 \ea
to separate the $\epsilon' \rightarrow 0$ term out from Eq.\
(\ref{self-energy_eff}) as
 \ba \label{a2fp0}
\Sigma_{eff}(\vk,\omega) =
 \sum_{\mathbf{k'}} G(\mathbf{k'},\omega) \alpha^{2}F_0(\mathbf{k,k'}),
\\ \alpha^{2}F_0(\mathbf{k,k'}) \equiv \lim_{f\rightarrow 0} s
\coth(s f/2T) \alpha^{2}F^{(+)}(\mathbf{k,k'},s f),
 \ea
where $s$ is the step size of numerical integration, and the
infinitesimal $\epsilon'\rightarrow 0$ is written as $\epsilon' =s
f$. Comparing Eq.\ (\ref{a2fp0}) with (\ref{self-energy_imp}) we
have
 \ba
n_{imp} \Big| V_{imp}(\mathbf{k},\mathbf{k'}) \Big|^{2} =
\alpha^{2}F_0(\mathbf{k,k'}).
 \ea
This means that if we take the zero frequency limit of the
Eliashberg function such that
 \ba \label{zero}
\lim_{f\rightarrow 0} \alpha^{2}F^{(+)}(\mathbf{k,k'},s f) =
\frac{f}{2T} n_{imp} \Big| V_{imp}(\mathbf{k},\mathbf{k'})
\Big|^{2},
 \ea
the impurity potential term may be expressed as the zero frequency
component of the boson spectrum. That is, the self-energy of Eq.\
(\ref{self-energy}) may be written as
 \ba \label{sigma2}
\Sigma(\mathbf{k},\omega) = \int^{\infty}_{-\infty}
d\varepsilon\int^{\infty}_{-\infty} d\varepsilon'
\frac{f(\varepsilon)+n(-\varepsilon')}{\varepsilon+\varepsilon'-\omega-i\delta}
 \sum_{\mathbf{k'}} A(\mathbf{k'},\varepsilon)
\alpha^{2}F^{(+)}(\mathbf{k,k'},\varepsilon'),
 \ea
together with Eq.\ (\ref{zero}).

In order to calculate the Eliashberg function $\alpha^{2}F^{(+)}$,
we perform the $k'_\perp$ summation of $\vk'$ using
 \ba
\int dk'_\perp A(\vk',\epsilon) =\frac{1}{\hbar v_F(\theta')} Re
N(\theta',\epsilon), \nonumber \\
 N(\theta',\epsilon) =
\frac{\epsilon}{\sqrt{\epsilon^2-\Delta^2(\theta',\epsilon)}},
 \ea
and rewrite Eqs.\ (\ref{sigma2}) as
 \ba \label{self_eff2}
\Sigma(\theta,\omega) = \int^{\infty}_{-\infty} d\epsilon'
M(\omega,\epsilon') \alpha^2 F^{(+)}(\theta,\epsilon'),
 \ea
where
 \ba
M(\omega,\epsilon')= \int^{\infty}_{-\infty} d\epsilon
\frac{f(\epsilon)+n(-\epsilon')}{\epsilon+\epsilon'-\omega-i\delta}
\left< Re N(\theta',\epsilon) \right>_{\theta'}, \\
 \alpha^2 F^{(+)}(\theta,\omega)
 = \left< \frac{\alpha^{2}(\theta,\theta')}{v_F(\theta')}
 F^{(+)}(\theta,\theta',\omega) \right>_{\theta'}.
 \ea
Then, using the MDC extracted self-energy as an input (the left
hand side of Eq.\ (\ref{self_eff2})), the Eliashberg function
$\alpha^2 F^{(+)}(\theta,\omega)$ may be straightforwardly
obtained by inverting this equation. The inversion was performed
using the maximum entropy method (MEM).

\vspace{1cm}

{\bf The self-energy and parameters of impurity potential \\ }

As explained in the main text the low energy feature in the ARPES
experiments in BSCCO was modeled by the off-plane impurity
potential. The fitting of the imaginary part of the self-energy
from UD89K with the impurity potential was presented in the main
text. We here show for completeness the plots of fitting the real
part of the self-energy of UD89K and the real and imaginary parts
of the self-energy from OD82K in Fig.\ 3.

The parameters obtained from the fitting of UD89K and OD82K are
listed in Table 1. $\Gamma(\theta)$ and $\Gamma_V (\theta)$ are
from Eqs.\ (7) and (12), respectively.

\begin{figure}[tbh]
\includegraphics[width=10.cm]{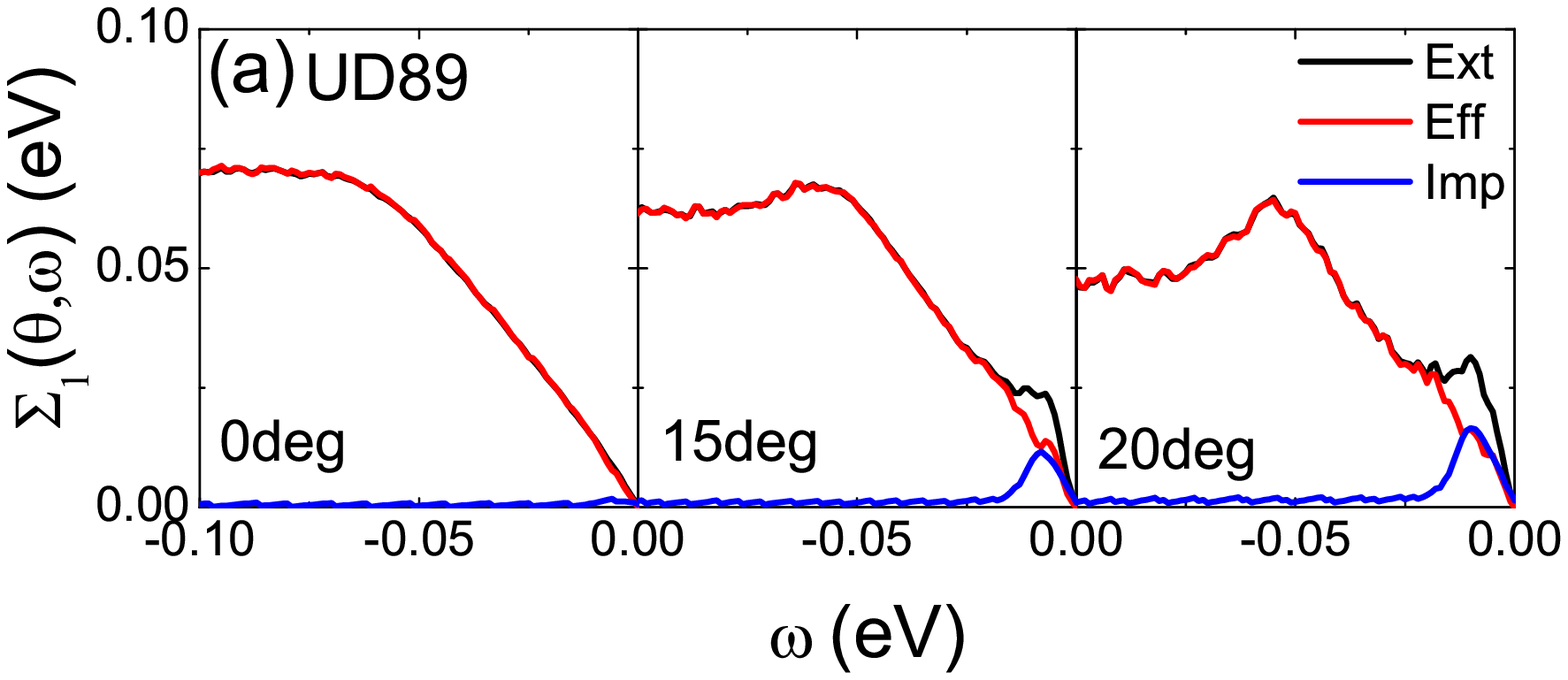}
\includegraphics[width=10.cm]{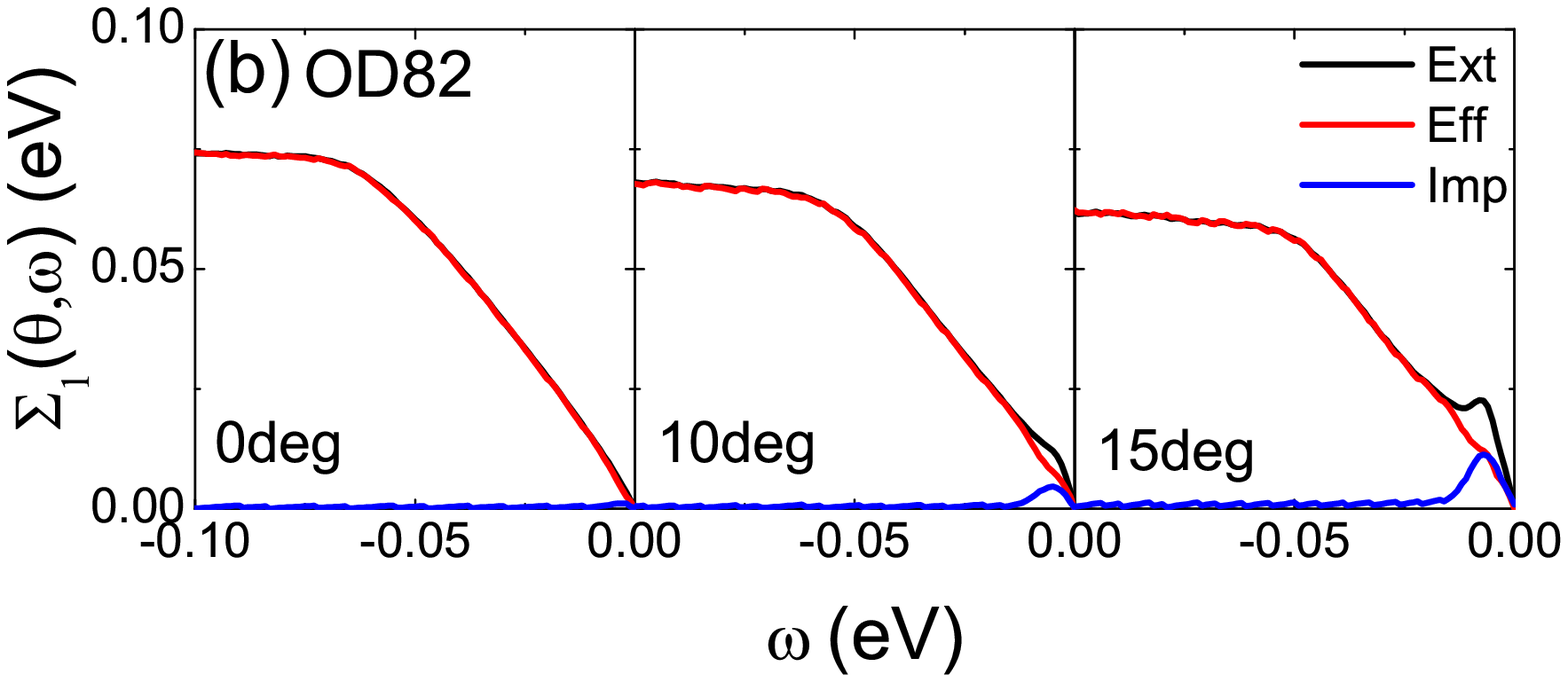}
\includegraphics[width=10.cm]{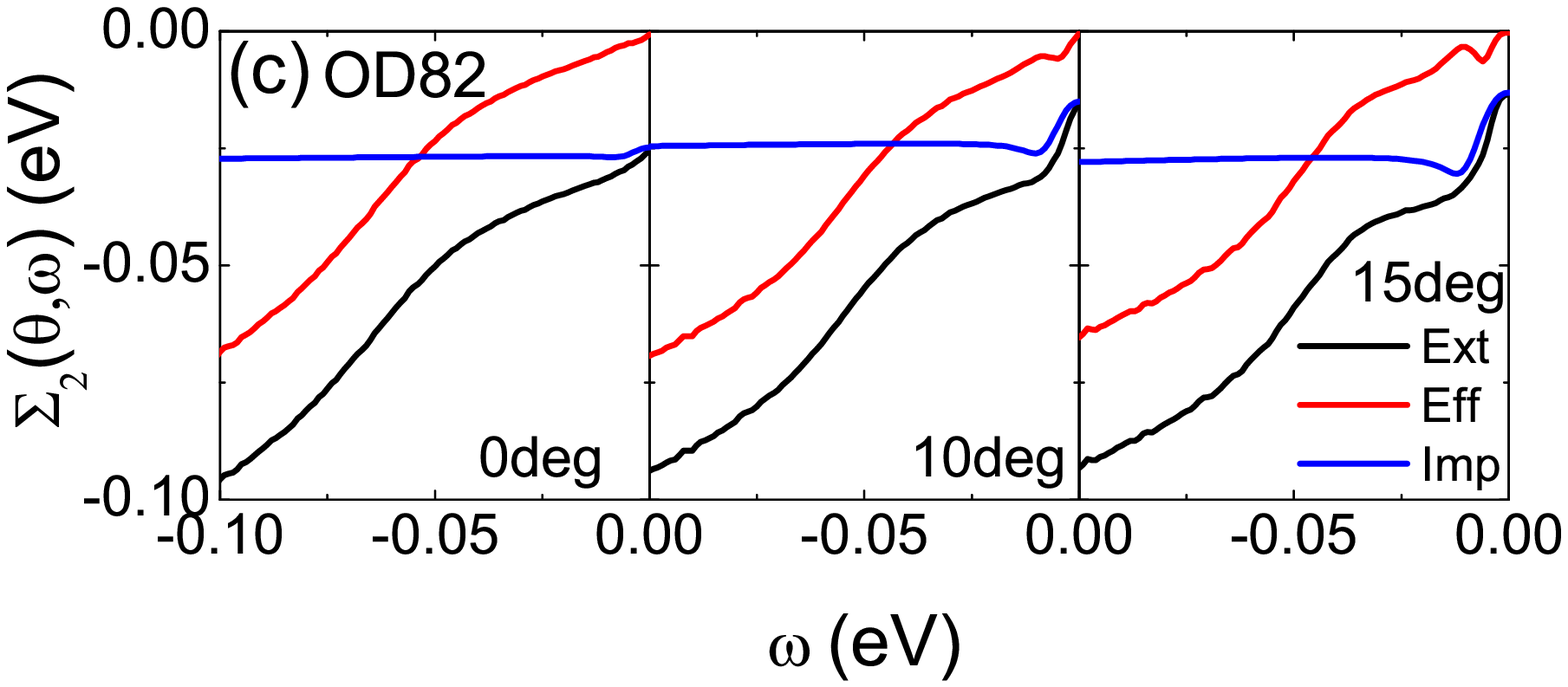}
\caption{ Comparison of the calculated impurity self-energy (blue
curves) with the MDC deduced one (black) for UD89K, as explained
in the main text. The red curves represent their difference, that
is, the $\Sigma_{eff}$ due to the coupling to bosons. Figure (a)
is the comparison of the real parts of UD89K at 16 K, and (b) \&
(c) are, respectively, comparison of the real and imaginary parts
of the OD82K at 17 K. }
 \label{fig:imp_re}
\end{figure}

\begin{table}[ht]
\caption{\label{tab:vimp} Parameters of the impurity potential
from fitting the self-energy of UD89K and OD82K as shown in Fig.\
4 of the main text and Fig.\ 3 of the SM. The $\Gamma(\theta)$ and
$\Gamma_V (\theta)$ are the estimates from Eqs.\ (7) and (12),
respectively. }
\begin{ruledtabular}
\begin{tabular}{l|lll|lll}
                    & \multicolumn{3}{l}{UD89K at 16 K}  & \multicolumn{3}{l}{OD82K at 17 K} \\
                    \hline
 $\theta$ (deg)                   & 0      & 15       & 20      & 0      & 10       & 15     \\
 \hline
 $\kappa~ (1/a)$                  & 0.3    & 0.3      & 0.3        & 0.3        & 0.3         & 0.3        \\
 $n_{imp}V_0^2 ~(t^2 a \kappa^3)$ & 0.03   & 0.033    & 0.03       & 0.045      & 0.035       & 0.035      \\
 $\Gamma_0$ (meV)                 & 1.5    & 2.5      & 3.5        & 2.5        & 3.0         & 3.3        \\
 $\phi_0$ (meV)                   & 20     & 20       & 18         & 16         & 16          & 16         \\
 $\Gamma$ (meV)            & 13     & 21       & 23         & 19         & 20          & 23         \\
 $\Gamma_{V}$ (meV)             & 17     & 23       & 24         & 26         & 22          & 24
\end{tabular}
\end{ruledtabular}
\end{table}

\bibliographystyle{apsrev}

\bibliography{review}